\documentclass[lettersize,journal]{IEEEtran}
\usepackage{amsmath,amsfonts}
\usepackage{algorithmic}
\usepackage{algorithm}
\usepackage{array}
\usepackage{enumitem}
\usepackage[caption=false,font=normalsize,labelfont=sf,textfont=sf]{subfig}
\usepackage{textcomp}
\usepackage{stfloats}
\usepackage{url}
\usepackage{verbatim}
\usepackage{graphicx}
\usepackage{hyperref}

\usepackage{tablefootnote}
\usepackage{scalerel}
\usepackage{tikz}
\usetikzlibrary{svg.path}
\definecolor{orcidlogocol}{HTML}{A6CE39}
\tikzset{
  orcidlogo/.pic={
    \fill[orcidlogocol] svg{M256,128c0,70.7-57.3,128-128,128C57.3,256,0,198.7,0,128C0,57.3,57.3,0,128,0C198.7,0,256,57.3,256,128z};
    \fill[white] svg{M86.3,186.2H70.9V79.1h15.4v48.4V186.2z}
                 svg{M108.9,79.1h41.6c39.6,0,57,28.3,57,53.6c0,27.5-21.5,53.6-56.8,53.6h-41.8V79.1z M124.3,172.4h24.5c34.9,0,42.9-26.5,42.9-39.7c0-21.5-13.7-39.7-43.7-39.7h-23.7V172.4z}
                 svg{M88.7,56.8c0,5.5-4.5,10.1-10.1,10.1c-5.6,0-10.1-4.6-10.1-10.1c0-5.6,4.5-10.1,10.1-10.1C84.2,46.7,88.7,51.3,88.7,56.8z};
  }
}

\newcommand\orcidicon[1]{\href{https://orcid.org/#1}{\mbox{\scalerel*{
\begin{tikzpicture}[yscale=-1,transform shape]
\pic{orcidlogo};
\end{tikzpicture}
}{|}}}}

\begin{document}

\title{Speaker Retrieval in the Wild:\\ Challenges, Effectiveness and Robustness}

\author{
    Erfan~Loweimi \orcidicon{0000-0002-8761-021X} ~\IEEEmembership{(Member,~IEEE)}\,,
    Mengjie~Qian \orcidicon{0000-0003-2614-5411} ~\IEEEmembership{(Member,~IEEE)}\,,
    Kate~Knill \orcidicon{0000-0003-1292-2769} ~\IEEEmembership{(Senior Member,~IEEE)}\,,
    Mark~Gales \orcidicon{0000-0002-5311-8219} ~\IEEEmembership{(Fellow,~IEEE)}
    
\thanks{Manuscript received dd mmmm yyyy; This is 1st place holder; This is 2nd place holder; This is 3rd place holder; This is the last place holder.}

\thanks{E. Loweimi (Corresponding author), M. Qian, K. Knill and M. Gales are with the Speech Group, Machine Intelligence Laboratory, University of Cambridge; e-mail: {\{el584, mq227, kmk1001, mjfg100\}@cam.ac.uk}.}

\thanks{E. Loweimi, M. Qian and M. Gales are supported by the EPSRC-funded Multimodal Video Search by Examples (MVSE) Project (EP/V006223/1). K. Knill is supported by the ALTA Institute, funded by Cambridge University Press \& Assessment (CUP\&A).}}

\maketitle

\begin{abstract}
There is a growing abundance of publicly available or company-owned audio/video archives, highlighting the increasing importance of efficient access to desired content and information retrieval from these archives.
This paper investigates the challenges, solutions, effectiveness, and robustness of speaker retrieval systems developed ``in the wild'', which involves addressing two primary challenges: extraction of task-relevant labels from limited metadata for system development and evaluation, as well as the unconstrained acoustic conditions encountered in the archive, ranging from quiet studios to adverse noisy environments. 
While we focus on the publicly available BBC Rewind archive (spanning 1948 to 1979), our framework addresses the broader issue of speaker retrieval on extensive and possibly aged archives with no control over the content and acoustic conditions.
Typically, these archives offer a brief and general file description, mostly inadequate for specific applications like speaker retrieval, and manual annotation of such large-scale archives is unfeasible. 
We explore various aspects of system development (e.g., speaker diarisation, embedding extraction, query selection) and analyse the challenges, possible solutions, and their functionality.
To evaluate the performance, we conduct systematic experiments in both clean setups and against various distortions simulating real-world applications. 
Our findings demonstrate the effectiveness and robustness of the developed speaker retrieval systems, establishing the versatility and scalability of the proposed framework for a wide range of applications beyond the BBC Rewind corpus.
\end{abstract}


\section{Introduction}
\IEEEPARstart{I}{n} the landscape of multimedia information retrieval \cite{stone2014multimedia,rueger-2010} and spoken document retrieval \cite{speaker_ret2023,content-ret-2015,content-ret-2012,Gokhan2011}, the demand for robust and effective speaker retrieval systems \cite{CAMPOS2019153,loweimi24_interspeech,qian2024zeroshot} is on the rise across various domains, such as multimedia indexing \cite{wagenpfeil2021ai}, surveillance and security \cite{security2022,Surveillance2014}, forensics \cite{forensics2023,PEDAPUDI2023100860,essery23_interspeech,forensics2022}, and copyright protection \cite{sturm2019ai,holzapfel2018ethical}. However, in real-world scenarios, the unpredictable nature of environments presents significant challenges to the development of efficient and reliable speaker retrieval systems. This challenge is particularly evident when dealing with extensive and potentially aged archives of media companies like the BBC, where recording conditions, technologies, and archiving practices have evolved over time. 
Consequently, innovative solutions are needed to tackle the complexities of speaker retrieval in dynamic and uncontrolled settings, where reliable task-specific labels or metadata for training and/or evaluation are often lacking. 
Moreover, such archive data encompass a wide range of acoustic conditions, spanning from high-quality studio recordings to noisy data captured in adverse real-world environments. These issues highlight the importance of the development of speaker retrieval systems for media archives that can effectively handle the conditions encountered in real-world applications.

Unlike \textit{speaker diarisation} which aims at answering ``who spoke when?'' \cite{dia-2022}, or \textit{speaker identification}, which focuses on determining the identity of a speaker within a closed-set \cite{Review2021}, or \textit{speaker verification}, which seeks to confirm whether a claimed identity matches a provided sample \cite{asv2022}, \textit{speaker retrieval} aims to find all instances of a target speaker across a large database or archive and rank them according to their relevance to the query speaker. 
While speaker identification primarily focuses on accuracy, speaker verification emphasises the equal error rate (EER) and speaker diarisation applies diarisation error rate (DER), in speaker retrieval the information retrieval performance metrics such as Precision@K are used for evaluation purposes.
Compared to other tasks, speaker retrieval is particularly challenging due to the need to efficiently search through extensive and potentially noisy archives, where the target speaker's recordings may vary significantly in terms of acoustic conditions, lingual content, emotional status, quality, intelligibility and length. As a result, a reliable speaker retrieval system must effectively handle a wide range of variations in speech signal characteristics to accurately locate the relevant documents.

The speaker retrieval workflow entails comparing a query clip with the archive files and ranking them based on relevance. Central to this process is the extraction of embeddings, ideally capturing the speaker's identity while remaining invariant to other information encoded in speech \cite{Bengio2013,Review2021}.
However, filtering out irrelevant attributes and generating embeddings solely for speaker discrimination poses a major challenge. It requires powerful models for representation learning along with extensive and diverse training data to enable the disentanglement of nuisance information while retaining only what is necessary for optimal speaker characterisation.
Development of various speaker embedding extraction methods such as i-vector \cite{ivector-2011}, d-vector \cite{dvector-2014,dvector2017}, x-vector \cite{xvector2018}, \cite{liu18b_interspeech}, phonetic-aware methods \cite{liu18b_interspeech,jin2023phoneticaware}, ECAPA-TDNN \cite{ecapa2020} and TitaNet (t-vector) \cite{TitaNet2021} models demonstrates efforts in this direction.

Another significant challenge is that archives typically provide only a brief and general content description, or synopsis, per file, which is often inadequate for extracting reliable task-specific labels.
For example, names mentioned in a synopsis might correspond to individuals who are visible but inaudible in the video (or vice versa), leading to a label appropriate for face retrieval but unreliable for speaker retrieval.
This indicates that the connection between synopsis-derived labels and the modalities is not always straightforward. That is, a label derived from a synopsis may exhibit an \textit{audio-visual presence}, or a \textit{modality-specific presence}, which can introduce noise and undermine the reliability of the labels in the task of interest.
Therefore, converting these synopses into reliable task-specific labels is a challenging endeavour, complicating their application for training and evaluation purposes.

In this article, the concept of \textit{development in the wild}, encompasses addressing two primary challenges. Firstly, it arises from the reliance on unreliable and noisy labels or metadata, a common predicament when dealing with real-world and aged archives where there may be lack of reliable task-relevant labels for system development and evaluation purposes. For instance, the metadata in our target database (BBC Rewind corpus \cite{bbcrewind}), despite containing valuable information briefly outlining events and people related to the video content, introduces ambiguity as it may include individuals with no voiceprint in the corresponding video. 

Secondly, \textit{development in the wild} also implies an unconstrained acoustic environment \cite{mclaren16_interspeech,voxceleb2017,voxceleb2020}. Our target archive covers data from 1948 to 1979, representing a substantial temporal span with varying technological and acoustic conditions, including diverse environments ranging from studio recordings to data recorded in the streets, demonstrations, etc. 
The speakers also exhibit a wide spectrum of emotional states, from calm and neutral to highly emotional and excited. 
Moreover, the database includes multi-talker scenarios and overlapping speech, which further complicates the analysis and processing of the data.
Such unconstrained conditions give rise to a great deal of intrinsic\footnote{Intrinsic factors relate inherently to the speaker him/herself and encompass variables like gender, age, dialect, accent, emotion, speaking style, etc.} and extrinsic\footnote{Extrinsic factors correspond to external influences, including reverberation, noise, and channel or microphone effects.} variations \cite{stoll2011}.

Addressing these challenges, we present a speaker retrieval framework that leverages state-of-the-art speaker diarisation and speaker embedding extraction models, trained on large-scale datasets. To this end, we will investigate the effectiveness of three widely-used embeddings, namely x-vector \cite{xvector2018}, ECAPA-TDNN \cite{ecapa2020} and TitaNet \cite{TitaNet2021} as well as their combination, and evaluate their performance across diverse scenarios. 
Moreover, we scrutinise the technical intricacies of system development, detailing the methodology, experimental setup, and results obtained from an extensive set of evaluations. The employment of various retrieval performance metrics, alongside examination of several types of distortion, including background noise, bit depth change, sampling rate mismatch and reverberation in both synthetic and real rooms, provide a thorough assessment of the systems' effectiveness and robustness. 

This work makes the following key contributions:	
\begin{itemize}
    \item Formulation of the problem of speaker retrieval in the wild on large-scale media archives, dealing with unreliable metadata and unconstrained acoustic conditions.
    
    \item Systematic analysis of metadata-derived label ambiguities, their implications for training and evaluation, and suggesting practical solutions.
    
    \item Extensive benchmarking of state-of-the-art speaker embeddings (x-vector, ECAPA-TDNN, TitaNet-Small and TitaNet-Large) under both clean and adverse conditions.
    
    \item Proposal of novel segment duration-based weighting strategies for speaker embedding aggregation.

    \item Comprehensive robustness evaluation, revealing model-specific strengths and weaknesses under diverse acoustic distortions, including additive background noise, reverberation, and mismatches in sampling rate and bit-depth.
    
\end{itemize}

The rest of this paper is organised as follows: Section 2 describes various components of the speaker retrieval system. Section 3 explores the BBC Rewind Archive and analyses it from various perspectives. Section 4 elucidates the system development in the wild, covering its what, how, and why. Section 5 is dedicated to presenting the experimental setup, results and discussion in a clean setup. Section 6 will explore the robustness of the system under various types of distortion and adverse acoustic conditions. Finally, Section 7 concludes the paper and outlines potential avenues for future work.

\section{Speaker Retrieval System}
\begin{figure}[!t]
\centering
  \includegraphics[width=\linewidth, height=37mm]{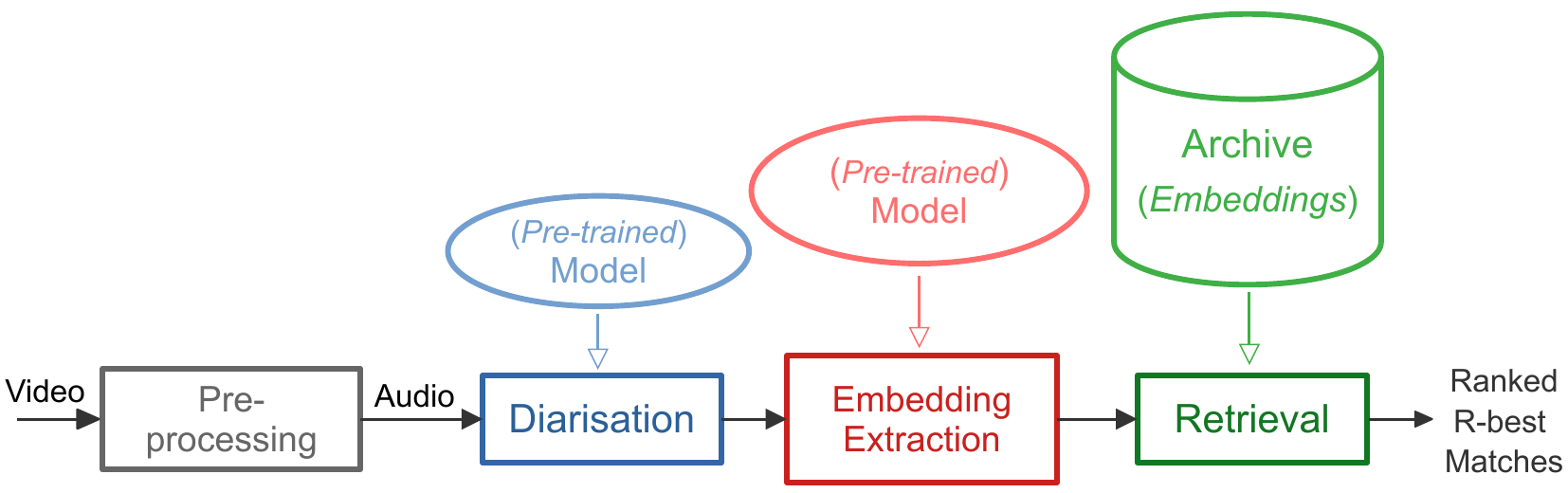}
  \caption{Workflow of the speaker retrieval system consisting of pre-processing, speaker diarisation, speaker embedding extraction and retrieval stages.}
  \label{fig:workflow}
\end{figure}

Fig.~\ref{fig:workflow} shows the workflow of our speaker retrieval system, consisting of pre-processing, speaker diarisation, speaker embedding extraction, and retrieval.
In this section, we provide a concise overview of each component.

\subsection{Pre-processing}
The initial step involves extracting audio from video files by transforming video (e.g., mp4) into audio (e.g., wav). The resulting audio file undergoes additional processing, encompassing waveform (volume) normalisation and sampling rate adjustments to match the target sampling rate. The target sampling rate is determined by the pre-trained models utilised in the downstream. In the pipeline, the first block after pre-processing is the diarisation; hence, the target sampling rate should match that of the diarisation model, which is 16 kHz.

\subsection{Speaker Diarisation}
The next block is speaker diarisation which aims at answering ``who spoke when?'' through temporal decomposition of the speech signal into distinct segments, each homogeneously corresponding to a specific speaker. 
The embedding extraction process relies on the diarisation results, and any inaccuracies in segmenting or assigning speakers can propagate through the pipeline, potentially affecting the quality of the extracted embeddings. To address this, we employ the widely used open-source PyAnnote.audio (ver 2.1) diarisation library \cite{Bredin2020,Bredin2021}, which is recognised for its state-of-the-art performance in speaker segmentation.
In the pre-trained PyAnnote diarisation pipeline, the hyperparameters of each block are optimised jointly and end-to-end to minimise the diarisation error rate (DER). Each block in the PyAnnote pipeline performs a sequence labeling task and is typically implemented using PyanNet-based architecture \cite{Bredin2020,Bredin2021} which is a cascade of convolutional, recurrent (Bi-directional LSTMs \cite{Schuster1997,Graves2005}) and fully-connected layers. 
Pyannote demonstrates (close to) state-of-the-art performance in many tasks, exhibits proficiency in handling overlapping speech, and eliminates the need to pre-determinedly specify the number of speakers.

\begin{figure}[t]
\centering
  \includegraphics[width=0.88\linewidth,height=100mm]{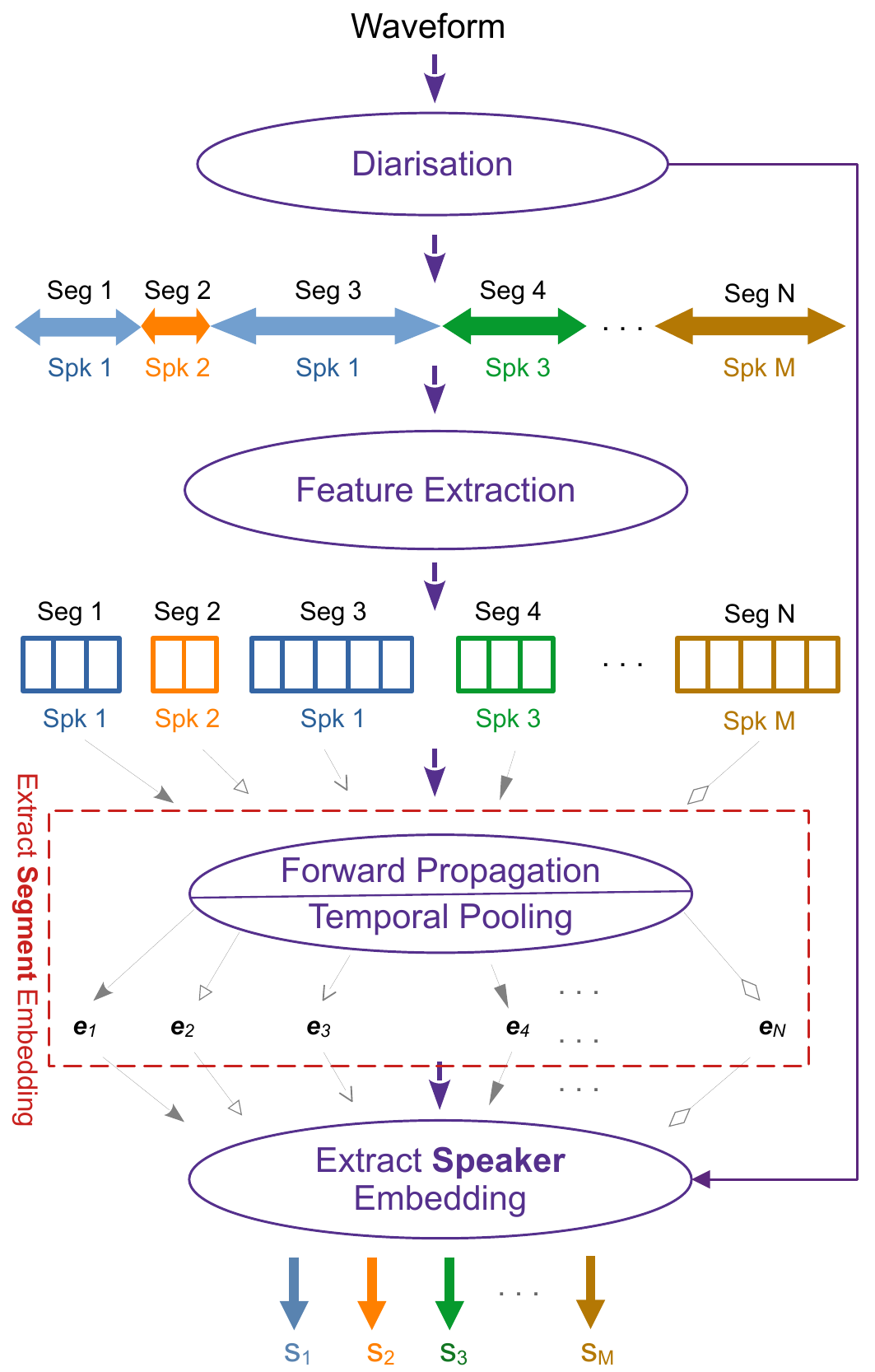}
  \caption{Extraction of segment ($\mathbf{e}$) and speaker ($\mathbf{s}$) level embeddings using pre-trained diarisation and embedding extraction models from input audio. $N$ and $M$, respectively, denote the number of segments and speakers, both automatically determined by the speaker diarisation block.}
  \label{fig:embed-ext}
\end{figure}

\subsection{Embedding Extraction}
\label{sec:embed}
Fig.~\ref{fig:embed-ext} shows the embedding extraction process using pre-trained diarisation and speaker embedding extraction models.

The diarisation step divides the signal into $N$ segments and assigns a speaker label to each segment from the estimated $M$ speakers in the audio. Then, each segment is passed through the feature extraction block (e.g., filterbank) and turned into a sequence of \textit{frame-level} acoustic feature vectors which are forward propagated through a pre-trained embedding extraction model. The temporal pooling operates over abstract representations extracted by the model and converts variable-length segment representations into a fixed-length embedding. 

After extracting the \textit{segment-level} embeddings $\mathbf{e}$, we compute the \textit{speaker-level} embeddings for all $M$ speakers in the recording \{$\mathbf{s}_1$, $\mathbf{s}_2$, $\cdots$, $\mathbf{s}_M$\} by averaging the embeddings of segments corresponding to each speaker label obtained from speaker diarisation:

\begin{align}
\mathbf{s}_i = \sum_{j=1}^{N} \ w_{ij} \ \mathbf{e}_j, \qquad \sum_{j=1}^N \ w_{ij} \ = \ 1, \qquad w_{ij} \ \geq \ 0. 
\label{eq:spk-emb}
\end{align}

\noindent where $N$, $\mathbf{s}_i$, $\mathbf{e}_j$ and $w_{ij}$ respectively indicate the number of segments, embedding of the $i^{th}$ speaker, embedding of the $j^{th}$ segment, and weight of the $j^{th}$ segment when computing the embedding of the $i^{th}$ speaker. When the $j^{th}$ segment does not belong to the $i^{th}$ speaker, $w_{ij}$ is zero.

The weighting process can take different forms. The simplest one is assigning identical weight to all segments belonging to one speaker, irrespective of segments' durations:

\begin{equation}
    w_{ij} = \frac{I_i(j)}{\sum_k \ I_i(k)} = \frac{I_i(j)}{N_i}
\end{equation}

\noindent where $N_i$ denotes the number of segments belonging to speaker $i$, and $I_i(j)$ is an indicator function, equal to one when segment $j$ belongs to speaker $i$, and is zero otherwise. Such a \textit{uniform} weighting scheme is suboptimal because shorter segments are typically less informative and may not adequately capture speaker's characteristics, whereas longer segments contain more information towards better characterising a speaker.

To address this issue, we make the weights proportional to the segments' length, assuming longer segments contain a larger speaker footprint and provide a more reliable characterisation for a speaker. There are multiple ways to make the weights proportional to the segment duration. For instance, the weights can be \textit{linearly} proportional to the duration

\begin{align}
    w_{ij} = \frac{\ I_{i}(j) \ d_j}{\sum_{k=1}^N I_{i}(k) \ d_k}.
    \label{eq:avg-lin}
\end{align}

\noindent where $d_j$ is the duration of the segment $j$. 

An alternative is the softmax function

\begin{equation}
    w_{ij} = \frac{I_{i}(j) \ \exp(d_j / \tau)}{\sum_{k=1}^N I_{i}(k) \ \exp( d_k / \tau)}
    \label{eq:avg-softmax}
\end{equation}

\noindent where $\tau$ denotes the temperature. This hyperparameter introduces flexibility into the weighting scheme: a higher temperature results in smoother weights (asymptotically approaching the uniform weights) while a lower temperature places a higher emphasis on longer segments.

There are other options such as using the \textit{ranks} of the segments belonging to each speaker

\begin{equation}
    w_{ij} = \frac{\textit{rank}(I_{i}(j) \ d_i)}{\sum_k \textit{rank}(I_{i}(k) \ d_i)} = \frac{\textit{rank}(I_{i}(j) \ d_i)}{\sum_{k=1}^{N_i} k} = \frac{\textit{rank}(I_{i}(j) \ d_i)}{N_{i} (N_{i}+1)/2}
    \label{eq:avg-rank}
\end{equation}

\noindent where \textit{rank} is an integer in the range of 1 to $N_i$, with the rank 1 and $N_i$ are respectively assigned to the shortest and longest segments of the $i^{th}$ speaker. Note that segments for which $I_{i}(j) = 0$ are excluded from the ranking process. 

In the ideal scenario, speaker embeddings should exclusively encapsulate the speaker's identity, and stay insensitive to various other information encoded in speech. 
This requires advanced representation learning techniques, ensuring that the embedding space is discriminative for speaker identity while being invariant to irrelevant variations. Here, we have employed the widely-used x-vector, ECAPA-TDNN and TitaNet models for speaker embedding extraction.

\subsubsection{x-vector}
The x-vector \cite{xvector2018} speaker embedding is obtained through a Time-Delay Neural Network (TDNN) \cite{tdnn1989}, trained for speaker identification tasks. The original architecture encompasses five 1-D convolutional layers, a statistical pooling layer and two fully-connected layers preceding a softmax speaker classifier. The time-delay convolution layers operate at the frame level. 
The statistical pooling layer calculates the temporal mean and standard deviation of the TDNN's frame-level outputs and effectively summarises a variable-length segment into a fixed-length vector. This vector is passed through the fully-connected layers towards the softmax classifier. All layers after the statistical pooling operate at the segment level.
Conventionally, the pre-activations of the first fully-connected layer are used as the x-vector embedding.

\subsubsection{ECAPA-TDNN}
The ECAPA-TDNN architecture stands out as a significant advancement in speaker embedding extraction. It draws inspirations from x-vector, sharing 1-D convolutions operating at the frame level along with statistical pooling and subsequent fully-connected layers. However, ECAPA-TDNN distinguishes itself via incorporation of four key blocks, namely SE-Res2Block \cite{ecapa2020}, multi-layer feature aggregation (MFA) \cite{ecapa2020}, attentive statistics pooling (ASP) \cite{asp2018} and additive angular margin (AAM) \cite{AAM2019,Xiang2019MarginMT} softmax. Both SE-Res2Block and MFA operate at the frame level, while layers subsequent to ASP work at the segment level.

The SE-Res2Block integrates a Res2Net backbone \cite{Res2Net2019} with a Squeeze-and-Excitation (SE) \cite{SE2018} block. MFA concatenated features at multiple levels of abstraction, right before extracting segment-level statistics. The ASP block employs a channel and context-dependent attention mechanism, where each frame and each channel is attended with a specific learnable weight. This dynamic and adaptive weighting scheme allows to focus on the most informative frames and channels. Finally, AAM enforces a higher intra-class similarity alongside a larger inter-class diversity.
These modifications collectively account for the superior performance of the ECAPA-TDNN over the x-vector.

\subsubsection{TitaNet}
TitaNet is inspired by ContextNet \cite{contextnet2020} and comprises an encoder with \textit{B} mega-blocks, each wrapped in a residual connection and containing \textit{R} repeats of certain sub-blocks. Each sub-block consists of a cascade of \textit{C} time-channel separable 1D convolutions \cite{Quartznet2020} (depthwise followed by pointwise convolutions), Batch Normalisation, ReLU activation, and Dropout. At the end of each mega-block, there is an SE block that extracts global contextual information and weights the local features accordingly. The decoder module consists of an ASP layer, followed by two linear layers and an AAM softmax. The output of the penultimate linear layer serves as the speaker embedding, referred to as the \textit{t-vector}. 

The configuration parameters B and R determine the depth, while C specifies the width of the network. In this study, we utilised two variants of TitaNet: TitaNet-Large (TitaNet-L) and TitaNet-Small (TitaNet-S). Both models have the same depth (B and R) but differ in width (C). For TitaNet-L, C is set to 1024, whereas for TitaNet-S, C is set to 256.

\subsection{Retrieval}
\label{sec:ret-spk-seg}
Having extracted the query embedding ($\mathbf{q}$), the score for each archive file (identified by \textit{fileID})  is computed as follows:

\begin{equation}
\text{score}[\textit{fileID}] = \max_{i \in 1, 2, \ldots, M_\textit{fileID}} \text{SimMetric}(\mathbf{q}, \mathbf{s}^{\textit{fileID}}_i)
\label{eq:ret-spk}
\end{equation}

\noindent where $M_\textit{fileID}$ represents the number of speakers in the \textit{fileID}, $\mathbf{s}^{\textit{fileID}}_i$ is the embedding of the $i^{th}$ speaker in the \textit{fileID} and the \textit{SimMetric} denotes a similarity metric, typically the cosine similarity. The \text{score} dictionary, with \textit{fileID} key and similarity score value, records the maximum similarity between the query and speakers in a file. Subsequently, the files are ranked based on their similarity score in descending order, and the top-R relevant files with the highest scores are retrieved.

It is worth noting that instead of speaker embeddings (\(\mathbf{s}_i\)), one can apply the segment embedding (\(\mathbf{e}_j\)):

\begin{equation}
\text{score}[\textit{fileID}] = \max_{j \in 1, 2, \ldots, N_\textit{fileID}} \text{SimMetric}(\mathbf{q}, \mathbf{e}^{\textit{fileID}}_j)
\label{eq:ret-seg}
\end{equation}

\noindent where $N_\textit{fileID}$ represents the number of segments in the \textit{fileID} and $\mathbf{e}^{\textit{fileID}}_j$ is the embedding of the $j^{th}$ segment.

In an ideal scenario where the embeddings effectively discriminate speakers according to their ID while being insensitive to nuisance and task-irrelevant information, the speaker-based and segment-based speaker retrievals should yield similar performance (assuming the segments are long enough to capture the speaker's characteristics). We will empirically compare these two approaches in Sections~\ref{sec:stats} and \ref{subsec:comp}.

Note that segment-level retrieval can be used to pinpoint specific segments and timestamps where the target speaker is present, offering a fine-grained analysis of speaker activity within a recording. In contrast, speaker-based retrieval provides a coarser granularity by determining whether the target speaker is ``present'' or ``not present'' in an entire recording, without identifying the precise segments of their activity. In this paper, our primary focus is on determining the presence of the target speaker in the recording.

\section{BBC Rewind Database}
\label{sec:rewind}
In this section, we investigate the properties of the BBC Rewind corpus \cite{bbcrewind} and analyse the provided metadata. 

\subsection{General Statistics}
\label{sec:stats}
The BBC Rewind dataset includes 43,059 videos and has four subsets: England, Scotland, Wales, and Northern Ireland. In the context of the EPSRC-funded MVSE project \cite{MVSE,MVSE-WU2024}, we focus on the Northern Ireland subset, although all the discussions and methods are equally applicable to other subsets. This subset comprises 12,594 video files, totaling 409 hours of content, and covers the period from 1948 to 1979. 

\begin{figure}[t!]
\centering
  \includegraphics[width=0.85\linewidth, height=95mm]{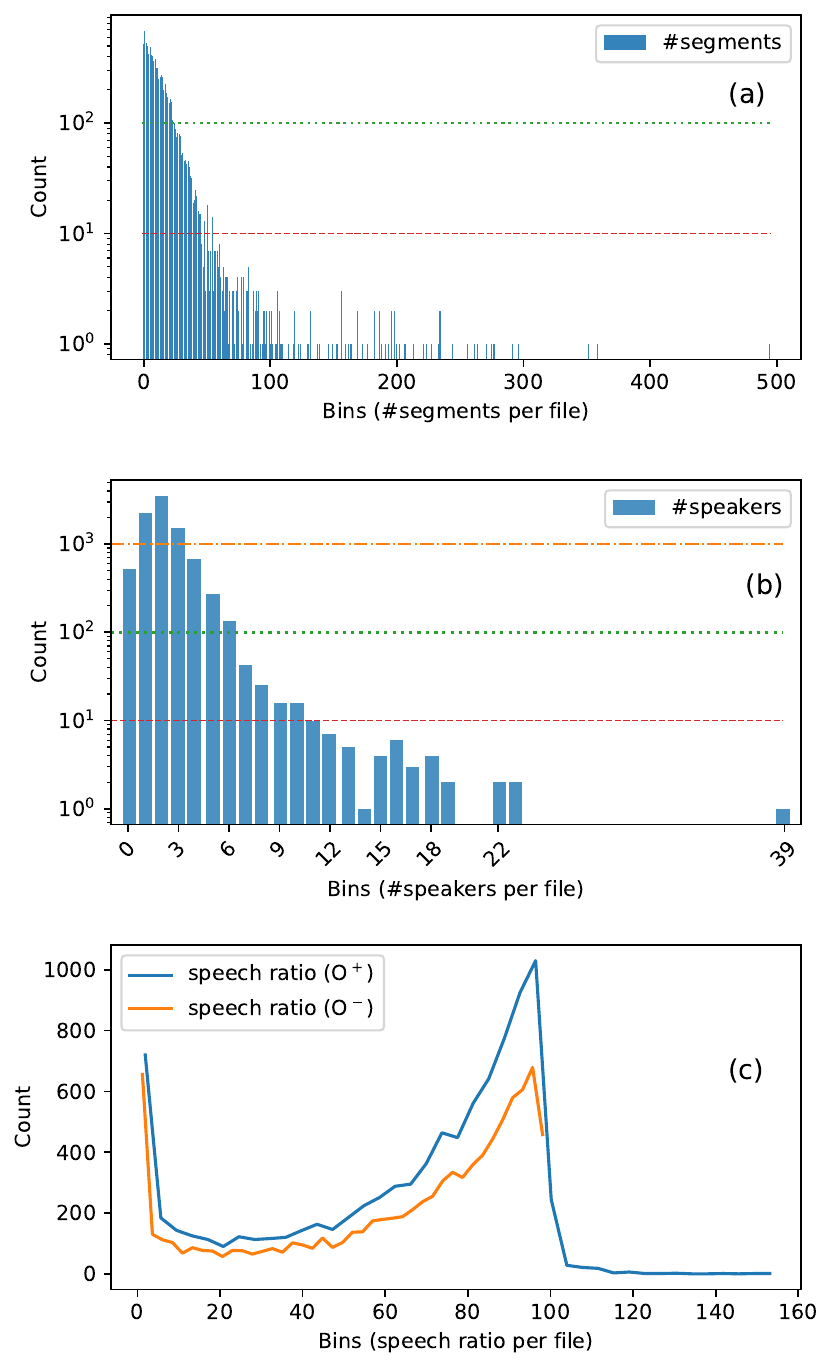}
  \caption{Histogram of the information extracted using diarisation. (a) number of segments (\#segments), (b) number of speakers (\#speakers), (c) speech ratio. The horizontal lines at 10, 100, and 1000 in (a) and (b) indicate levels to facilitate readability and comparison.}
  \label{fig:hist-dia}
\end{figure}

Table~\ref{tab:stats} reports the statistics (mean, median, standard deviation (STD), min and max) and Figs.~\ref{fig:hist-dia}, \ref{fig:hist-spk} and \ref{fig:hist-names} illustrate the histograms of various factors. 
As indicated in Table~\ref{tab:stats}, the mean and median of video durations are 117 and 83 seconds, respectively. Notably, the standard deviation is quite large (167 seconds), with the shortest video file being 3 seconds and the longest extending to 3,039 seconds.

The diarisation block provides useful information about each file such as number of segments (\#segments), number of speakers (\#speakers), and \textit{speech ratio}\footnote{speech ratio = speech\_duration / total\_video\_duration $\times$ 100}. Fig.~\ref{fig:hist-dia}~(a), (b) and (c) illustrate the corresponding histograms, and Table~\ref{tab:stats} presents the statistics. On average, files have two speakers, with the minimum of 0 and maximum of 39 speakers. Predominantly, files tend to have one to three speakers, with a mean and median of number of segments per file being 14 and 10, respectively. 
Thus, on average, the number of segments per file is approximately five to seven times larger. 
Additionally, based on diarisation, this archive includes 124,266 segments (or utterances) and 20,299 speakers\footnote{This is the sum of speakers detected per file by the diarisation, and does not mean there are 20,299 \textit{distinct} speakers in the archive.}. These observations suggest that speaker-based retrieval is expected to be approximately five to seven times more computationally efficient than segment-based retrieval, as it involves fewer comparisons and cosine similarity scores computation. 

The speech ratio, which represents the fraction of the speech parts to the total video duration, can be computed with and without considering the multi-talker overlaps. We refer to the former, $O^+$ and to the latter $O^-$. In $O^+$ segments duration is summed regardless of overlap while $O^-$ means the overlapping duration is counted only once, regardless of how many segments it spans. The difference between $O^+$ and $O^-$ reflects the contribution of overlapping speech and helps quantify its significance. Fig.~\ref{fig:hist-dia}~(c) and Table~\ref{tab:stats}, show the distribution and the statistics of speech ratio with and without overlap. As seen, for $O^+$, the speech ratio may exceed 100\%, owing to the presence of overlapping speech.
The overlap speech further complicates the speaker retrieval task because it leads to mixing up the features of different speakers, making the embeddings less accurate in representing each speaker.

Table~\ref{tab:stats} also shows the statistics of the signal-to-noise ratio (SNR). The SNR is estimated using the non-intrusive \textit{waveform amplitude distribution analysis} (WADA) method, proposed in \cite{kim2008}. The SNR statistics indicate a moderate level of noise across the dataset. The mean is 19.2 dB, with a median of 17 dB and a standard deviation of 10 dB. The range of SNR values spans from -8 dB to 60 dB, reflecting a diverse set of acoustic conditions encountered in the dataset.


Building on these statistics, the BBC Rewind corpus presents several practical challenges that highlight the complexities of working with real-world, uncontrolled audio data. The recordings encompass diverse acoustic conditions, ranging from studio-quality audio to highly noisy environments such as street interviews, public demonstrations, and crowded spaces, introducing significant background noise\footnote{Quantifying the specific noise types (e.g., traffic, crowd chatter, wind) and their levels in the BBC Rewind corpus is challenging, if not impossible, as this information is not included in the metadata, which was curated by journalists with a focus on content rather than acoustic properties.}. Also, the corpus includes recordings with speakers in various emotional states such as calm, neutral, excited, or distressed, which adds variability to speech characteristics and complicates speaker identification. Moreover, the presence of overlapping speech and variations in recording setups, including close and distant microphone placements, further increase the difficulty of extracting representative speaker embeddings. These challenges, combined with the natural variability in segment lengths and the diversity of topics covered (e.g., politics, health, education, science, business and sports), make the BBC Rewind corpus an ideal but demanding testbed for developing speaker retrieval systems.



\begin{table}[t]
\centering
\caption{{\it Statistics of various properties of the BBC Rewind corpus. $^{\dagger}$: Computed using speaker diarisation; $^{\ddagger}$: computed using metadata (synopses)}.}
\vspace{-3mm}
\begin{tabular}{l|ccccc}
\hline \\[-3.0mm]
 & Mean & Median & STD & Min & Max \\
\hline
\hline
\\[-3mm]
Video Duration (sec)  & 117 & 83 & 164 & 3  & 3039 \\
\hline
$^{\dagger}$Speech Ratio ($O^+$) & 66  & 77 & 31.8 & 0 & 155 \\
$^{\dagger}$Speech Ratio ($O^-$) & 65  & 76 & 30.8 & 0 &  99.3 \\
$^{\dagger}$\#segments        & 14  & 10 & 21  &  0 & 495 \\
$^{\dagger}$\#speakers        &  2.2&  2 &  1.7&  0 &  39 \\
$^{\dagger}$Segment - Duration (sec) & 8.1  & 4.5 & 10.1  &  0.2 & 241 \\
\hline
$^{\ddagger}$\#names           &  2.1 &  2 &  1.3 & 0 &  30  \\
$^{\ddagger}$\#words           & 20.4 & 17 & 13.5 & 1 & 324 \\ 
$^{\ddagger}$\#sentences       &  1.6 &  1 &  0.9 & 1 &  14  \\ 
\hline
SNR (dB)          &  19.2 & 17.0 &  10.3 & -8 &  60  \\ 
\hline
\hline
\end{tabular}
\label{tab:stats}
\end{table}

\subsection{Metadata}
Each video is accompanied by a metadata JSON file that offers general details, including location, time, and notably, a \textit{synopsis}. These synopses, curated by journalists, typically encapsulate the content in one or two sentences and crucially include the names of significant individuals, often politicians, featured in the video. We presume that these identified names correspond to the speakers in the video, and will examine the validity of this assumption in subsequent discussions.

To extract individuals' name from the synopses, we employed spaCy \cite{spacy} library wrapping the state-of-the-art transformer-based language models (\texttt{en\_core\_web\_trf}), leveraging its named entity recognition (NER) capabilities. 
To improve the accuracy of NER, a series of text normalisation steps were implemented to standardise the text. Initially, BeautifulSoup \cite{beautifulsoup} removed HTML tags, followed by the unicodedata \cite{unicodedata} module for consistent character representation. 
The NER process leads to extracting about 5,800 distinct names. 

\begin{figure}[t!]
\centering
  \includegraphics[width=0.95\linewidth,height=58mm]{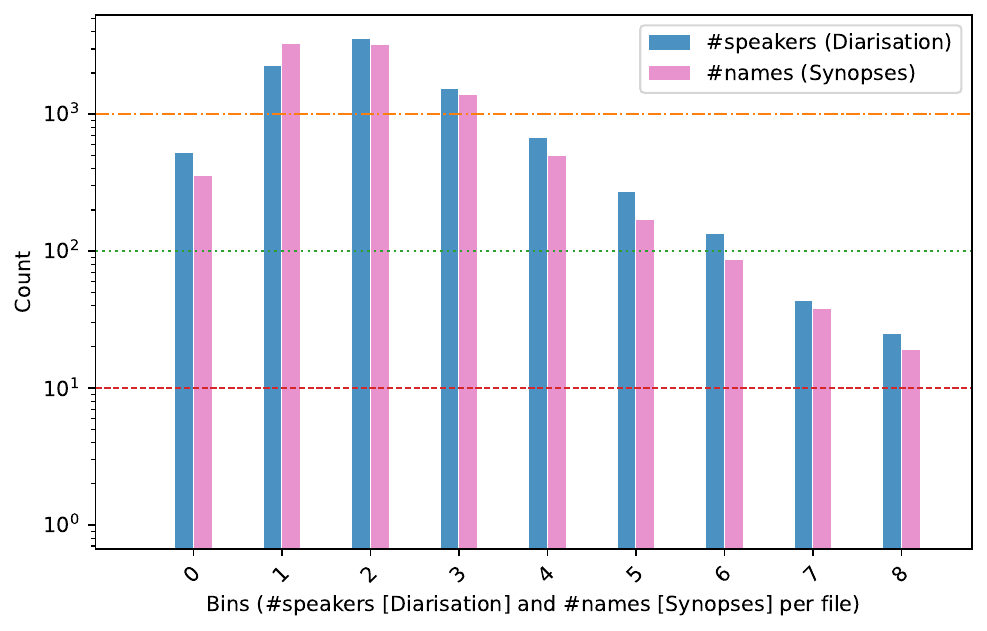}
  \caption{Histograms of the \#speakers per file (according to diarisation) along with the histogram of the \#names (according to synopses) per file.}
  \label{fig:hist-spk}
\end{figure}

Table~\ref{tab:stats} presents the statistics of the number of names (\#names) per file and Fig.~\ref{fig:hist-spk} juxtaposes the histograms of the \#names and \#speakers per file. The average \#names and \#speakers per file are 2.1 and 2.2, respectively, and both have a median of 2 as well as similar histograms. For example, only files with one to three speakers/names occur more than 1,000 times and only files with zero to five speaker/names have more than 100 occurrences.
While these similarities do not ensure a perfect alignment between \#names and \#speakers, they provide substantial evidence supporting the assumption of a reasonable correspondence.

\begin{figure}[t!]
\centering
  \includegraphics[width=0.95\linewidth,height=58mm]{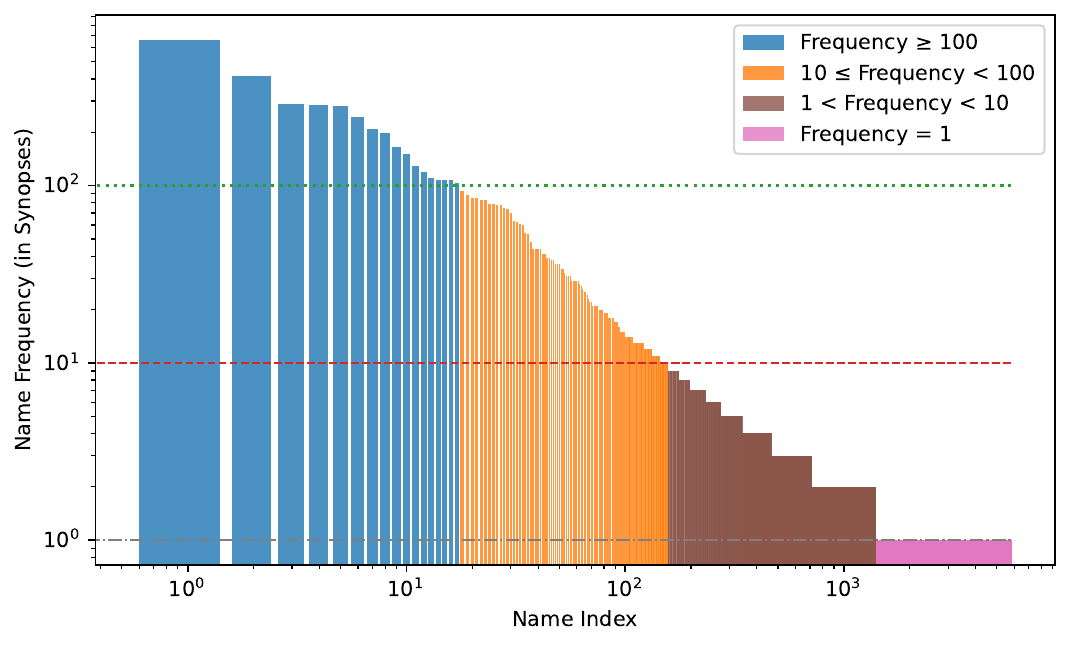}
  \caption{Occurrences of names vs.~name index, computed by calculating the frequency of individual names extracted from Rewind synopses by NER module. In total, NER detected over 5,800 distinct person names.
  }
  \label{fig:hist-names}
\end{figure}

Fig.~\ref{fig:hist-names} depicts the counts of the distinct names. The top three most frequent names occur 685, 429, and 301 times, respectively. Additionally, 20 names appear more than 100 times, while 140 names have appeared between 10 to 100 times. Furthermore, 1,240 names have occurred between 1 to 10 times, and 4,400 names appear only once. 

It is difficult, if not impossible, to estimate the exact number of speakers across the archive. However, assuming each name corresponds to a speaker, there would be approximately 5,800 speakers in the database. This presents a formidable challenge in accurately discriminating speakers and retrieving the desired one. That is, there are approximately 5,800 distinguished speaker embeddings, with some closely located to the target speaker. 
Furthermore, it is likely that the actual number of speakers exceeds the number of names, as the listed names in a synopsis may not encompass all individuals present in the videos, but rather highlight the most significant people. This inference is supported by Fig.~\ref{fig:hist-spk}, where typically \#speakers are slightly larger than \#names per bin.

\section{Development in the Wild}
\label{sec:wild}
As mentioned in Section~I, the concept of ``development in the wild'' in this paper, is prompted by the unconstrained environment and the lack of reliable labels. In this section we scrutinised the utility of the synopses-derived labels (speaker names) for training and performance evaluation purposes.

\subsection{Synopses Types and Nature of Videos}
\label{sec:wild-type}
After a meticulous analysis of the synopses and the corresponding video contents, we identified three scenarios that capture the relationship between the names mentioned in the synopses and the speakers heard in the videos:

\begin{itemize}
    \item \textit{Audio-Visual Presence (AVP)}: names correspond to individuals who are seen and heard in the video.

 
    \item \textit{Silent Presence (SP)}: names belong to individuals who have no corresponding voiceprint in the video.

    \item \textit{Audio-only Presence} (AoP): An individual mentioned in the synopsis is heard but is invisible in the video footage.

\end{itemize}

\noindent To further clarify these scenarios, let's look at some examples:

\begin{enumerate} 
    \item Audio-Visual Presence: \textit{\underline{Eric Waugh} interviews Prime Minister \underline{Brian Faulkner} on the Wilson Plan. He also interviews \underline{William Craig} and \underline{Ian Paisley}}; All names are present in both audio and video.


    
    \item Silent Presence: \textit{William Whitelaw is interviewed after \underline{Bernadette Devlin}, MP for Mid-Ulster, assaulted \underline{Reginald Maudling}, Home Secretary, in the House of Commons}; Both Bernadette Devlin and Reginald Maudling belong to the Silent Presence category.


    \item Audio-only Presence: \textit{Chief Superintendent Meharg is interviewed by \underline{Don Anderson} about precautions against crime}; Don Anderson is heard but is not seen.

\end{enumerate}


The provided examples illustrate the complexities involved when relying on names as surrogates for speakers. Although data from instances showcasing AVP and AoP can be reliably utilised for speech-based speaker retrieval, the SP scenario introduces ambiguity and inaccuracies into the process of speaker labelling using the names mentioned in synopses. Furthermore, accurately delineating the connection between mentioned names and actual speakers in the videos is intricate, as evidenced by the SP scenarios. In fact, without a thorough examination of the auditory content of the videos, solely relying on the synopses makes it uncertain whether individuals actually contribute vocally in the video.

\subsection{Training in the Wild}
\label{sec:wild-train}
Utilising video synopses to train models for speaker identification and retrieval encounters two serious challenges: noisy labels as well as ambiguous speaker-name associations. The presence of noisy labels is almost inevitable, as discussed earlier. 
In addition, establishing a clear mapping between synopses-derived names and the diarisation-derived speakers poses a significant problem. Without explicit cues from the video or additional metadata, accurately matching each name to the rightful diarisation speaker index becomes daunting, even when the number of names extracted from synopses matches the number of speakers identified by diarisation.

These two issues substantially undermine the utility of such data for training or fine-tuning purposes. 
In response to these challenges, leveraging pre-trained models in a \textit{zero-shot} approach emerges as a viable solution. However, this introduces another nuanced issue: the potential domain mismatch between the data on which these models were initially trained and the target BBC Rewind archive, which encompasses a wide array of acoustic conditions. Such diversity in acoustic conditions, along with potential discrepancies across other dimensions, can impact the effectiveness of the pre-trained models.

To enhance adaptability to real-world scenarios and successfully alleviate the domain mismatch, it's crucial to enrich and maximise the variability of the training data. This involves incorporating a diverse set of speakers, accents, acoustic conditions and background noise into the training dataset. 
These will enhance the model's generalisability and hopefully enable it to effectively handle diverse and unpredictable scenarios.

\subsection{Evaluation in the Wild}
\label{sec:wild-eval}
Despite the challenges associated with using synopses-derived labels for training, they possess significant value for performance evaluation. This is owing to the fact that in the evaluation phase, the speaker-name mapping challenge is relaxed. Essentially, if the query's speaker name appears among the list of names extracted from the synopsis of a retrieved file, it can be considered a valid hit (true positive).

Basically, two potential errors could compromise the reliability of the labels for evaluation purposes: either a name is erroneously missed or erroneously added 
to the synopsis. The probability of occurrence of both errors directly depends on the prominence of the query speaker. 
The greater the significance of the speakers, the less likely they are to be overlooked by the expert archivists when composing the synopses.
In fact, one of the primary objectives of the expert archivist who crafted the synopses is to include key figures associated with or appearing in the video. Given that our task focuses on retrieving speakers who are famous politicians in the UK and Northern Ireland, the likelihood of erroneously adding them to the synopses or missing them is minimal. Therefore, these labels can be safely applied for evaluation purposes.

\section{Exploring the Effectiveness}
\label{sec:exp}
\subsection{Experimental Setup}
To extract speaker embeddings, we applied the x-vector, ECAPA-TDNN and TitaNet models. For the x-vector we used SpeechBrain's implementation \cite{speechbrain_xvector}, for ECAPA-TDNN we utilised both the SpeechBrain's implementation (ECAPA-TDNN-SB) \cite{speechbrain_ecapa_tdnn} and NeMo's (ECAPA-TDNN-NeMo) \cite{nvidia_ecapa_tdnn} implementation, and for the TitaNet \cite{nvidia_titanet_l}, the NeMo toolkit was employed. The SpeechBrain models are trained solely on the VoxCeleb-1 \cite{voxceleb2017} and VoxCeleb-2 \cite{voxceleb2018} datasets which together feature over 2,000 hours data with 7,205 speakers. 
The VoxCeleb database is greatly suitable for our purpose of system development in the wild because: 
\begin{itemize}
    \item It is collected ``in the wild'', meaning that the speech segments are naturally contaminated with real-world noise, including laughter, cross-talk, channel effects, music, and other environmental sounds \cite{voxceleb2020};
    
    \item Furthermore, the dataset has a remarkable diversity due to its multilingual nature, featuring speech from speakers of 145 different nationalities, spanning a wide range of accents, ages, ethnicities, and languages \cite{voxceleb2020}.
\end{itemize}

The NeMo's models are trained with the VoxCeleb 1 and 2 as well as LibriSpeech \cite{Librispeech2015}, Switchboard \cite{SWITCHBOARD1992}, Fisher \cite{Fisher2004}, SRE \cite{sre:96-01} and RIR noise \cite{RIR-Noise-Database2017} databases. In total, this extensive training set features 3,373 hours of data and 16,681 speakers.

To assess the retrieval performance, we employ the widely-used Precision@K (P@K) metric: the number of relevant files within the top-K retrievals. Specifically, we calculate P@1, P@3, P@5 and P@10. A drawback of P@K is its disregard for the position of relevant items. However, this limitation diminishes when one simultaneously considers precision at multiple cutoff points such as P@1, P@3, P@5, and P@10. Additionally, we use Mean Average Precision@K (MAP@K), Mean Reciprocal Ranking (MRR) \cite{MRR}, and Normalised Discounted Cumulative Gain@K (NDCG@K) \cite{ndcg2002,ndcg2013} to evaluate the retrieval performance. In all these metrics, higher values indicate better retrieval performance. For more detail about the metrics, please refer to \cite{manning2009evaluation,Carterette2011}.


Our query set consists of 523 video files, totaling 21.1 hours of data, randomly selected from the Rewind corpus. These files represent 38 renowned politicians from Northern Ireland. 

We refer to the experiments in this section as the ``clean setup'', aimed at investigating the \textit{effectiveness} of the system. In Section~\ref{sec:robustness}, we contaminate the queries with various types of distortions which simulate adverse conditions, and analyse the system's \textit{robustness} in a ``noisy setup''.

\subsection{Initial Experimental Results}
As outlined in Section~\ref{sec:embed}, several methods exist for deriving the speaker-level embeddings from the segment-level ones. Table~\ref{tab:spk-avg} presents the retrieval performance of various averaging techniques, for the ECAPA-TDNN-SB model. Making the weights proportional to the segment duration proves beneficial. As discussed in Section~\ref{sec:embed}, shorter segments may not effectively characterise a speaker, so they should receive a smaller weight compared to longer segments. Among the methods tested, linear weighting demonstrates the highest performance, surpassing both softmax and ranking methods, without the need to adjust a hyperparameter, as required by the softmax method. 
Henceforth, we will apply the linear weighting method to compute the speaker-level embeddings.

\begin{table}[t]
\centering
\caption{{\it Speaker retrieval performance for various averaging methods used to extract speaker-level embeddings. The ECAPA-TDNN-SB embeddings are used here.}}
\label{tab:spk-avg}
\begin{tabular}{l|c|c|c|c}
\hline \\[-3mm]
 Method & P@1 & P@3 & P@5 & P@10 \\
\hline
\hline \\[-3mm]
Uniform & 70.9 & 69.5 & 67.3 & 63.7 \\
Linear & \textbf{74.6} & \textbf{72.7} & \textbf{70.5} & \textbf{66.9} \\
Softmax ($\tau=1$)  & 73.4 & 70.9 & 69.1 & 65.2 \\
Softmax ($\tau=5$)  & 73.6 & 71.6 & 69.9 & 65.9 \\
Softmax ($\tau=10$) & 74.4 & 72.3 & 70.3 & 66.3 \\
Softmax ($\tau=15$) & 73.8 & 72.0 & 70.2 & 66.3 \\
Ranking & 73.8 & 71.6 & 69.9 & 66.0 \\
\hline
\hline
\end{tabular}
\end{table}

\subsection{Performance Evaluation with Noisy Labels}
As discussed in Section~\ref{sec:wild-type}, the speaker labels are rather noisy, and only files with AVP and AoP synopsis types include the target speaker's voiceprint. To scrutinise this issue, we conducted a \textit{cumulative analysis} described below:

\begin{itemize}
    \item Sort all query files based on their performance:
    \begin{equation}
        Q^{\text{*}}_{1:N_Q} = \{\text{File}_1, \text{File}_2, ..., \text{File}_{N_Q}\}
    \end{equation}
    
    \noindent where $Q^{*}_{1:N_Q}$ is sorted query set with $N_Q$ query files (523, here), $\text{File}_1$ has the best performance, and $\text{File}_{N_Q}$ has the worst. By performance, we mean the average of the P@1, P@3, P@5 and P@10.

    \item Build a partial query set, $Q^*_{1:n}$, including the top-n files:
    \begin{equation}
        Q^*_{1:n} = \{\text{File}_1, \text{File}_2, ..., \text{File}_{n}\}
    \end{equation}

    \item Vary $n$ from 1 to $N_Q$ and compute the performance for $Q^*_{1:n}$. The y-coordinate on the cumulative performance graph represents Precision@K for $Q^*_{1:n}$ query set, while the x-coordinate corresponds to the cumulative query set coverage. This coverage is calculated in percent as the cardinality of $Q^*_{1:n}$, namely $n$, divided by $N_Q$.

\end{itemize}

The cumulative performance declines as the cumulative coverage of the query set increases. It is anticipated that the query files with noisy labels, specifically those falling into the \textit{Silent Presence} category, will be among the last files added to the sorted query set $Q^{*}$, because the retrieval performance for them is very poor. Appending the $Q^{*}$ with these type of queries will result in a consistent and noticeably sharp drop in the cumulative performance.

\begin{figure}[t!]
\centering
  \includegraphics[width=0.90\linewidth]{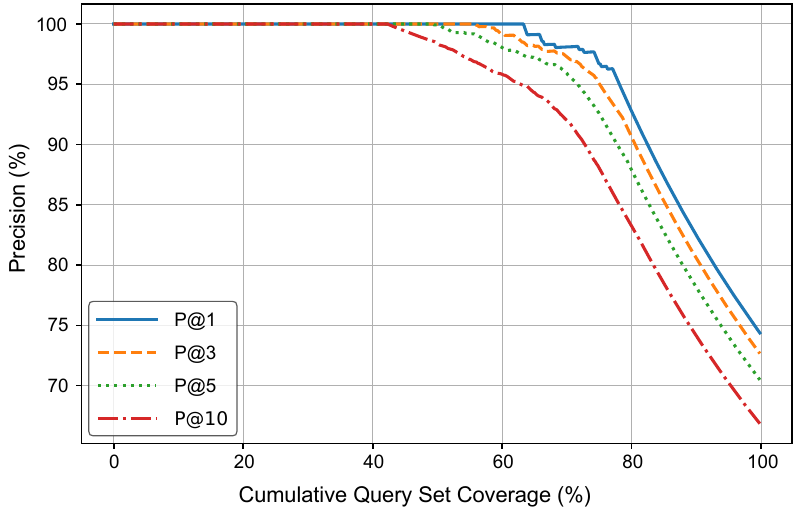}
  \caption{Cumulative performance analysis. Precision@K (P@\{1,3,5,10,\}) vs cumulative query set coverage for the ECAPA-TDNN-SB system.}
  \label{fig:cum-prec}
\end{figure}

Fig.~\ref{fig:cum-prec} displays the cumulative performance vs cumulative coverage. 
For about 60\% of the data, the performance is near-perfect in terms of P@1. For cumulative coverage between 60\% to 75\%, there is a performance drop, but it is not sharp. However, for cumulative coverage from 75\% to 100\%, we observe a sharp and consistent drop in P@1. This performance decline can be primarily attributed to the query files categorised as Silent Presence or to queries with low quality.



To gain a more precise understanding of the queries and their corresponding synopses types, we thoroughly examined both the visual and auditory contents of the videos in the query set. This analysis helped us determine the types of synopses associated with each query.
Table~\ref{tab:query-cat} presents \#Files, percentage and performance for each category. As observed, the performance for the SP scenario (which constitutes 15\% of the query set) is notably poor, whereas for the AVP and AoP categories, the performance is remarkably better. 


Henceforth, for performance evaluation, we use the union of the AVP and AoP categories (AVP $\cup$ AoP) as the query set and will refer to it as Q$^\dagger$. It includes 451 files and 19.3 hours data. 
This query set refinement mitigates concerns about attributing errors to noisy labels and enables a more accurate speaker retrieval performance assessment.

\begin{table}[t]
\centering
\caption{{\it Speaker retrieval performance for ECAPA-TDNN-SB speaker embeddings.}}
\label{tab:query-cat}
\begin{tabular}{l|c|c|c|c|c|c}
\hline \\[-3mm]
  Query & \#Files & Hours & P@1 & P@3 & P@5 & P@10 \\
\hline
\hline \\[-3mm]
 All     & 523 & 21.1 & 74.6 & 72.7 & 70.5 & 66.9 \\
 \hline
 AVP     & 425 & 18.3 & 86.6 & 84.3 & 82.2 & 78.3 \\
  SP     &  72 &  1.8 &    0 & 3.7  & 3.1  & 2.6  \\
 AoP     &  26 &  1.0 & 80.8 & 73.1 & 64.6 & 57.7 \\
 \hline
 AVP $\cup$ AoP & 451 & 19.3 & 86.3 & 83.7 & 81.2 & 77.1 \\
\hline
\hline
\end{tabular}
\end{table}

\begin{table}[t]
\centering
\caption{{\it Retrieval performance for various speaker embeddings. XV: x-vector, ET: ECAPA-TDNN, SB: SpeechBrain, TS: TitaNet-Small, TL: TitaNet-Large.}}
\label{tab:metrics}
\begin{tabular}{l|c|c|c|c|c}
\hline \\[-3mm]
        & XV-SB & ET-SB & ET-NeMo & TS-NeMo & TL-NeMo \\
\hline   
\hline
P@1     & 73.4 & 86.3 & 84.9 & 86.5 & \textbf{86.9} \\
P@3     & 66.8 & \textbf{83.7} & 82.4 & 83.4 & 83.3 \\
P@5     & 63.5 & \textbf{81.2} & 80.8 & 80.7 & 81.1 \\
P@10    & 54.7 & \textbf{77.1} & 75.8 & 75.3 & 76.2 \\
\hline
MAP@1   & 73.4 & 86.3 & 84.9 & 86.5 & \textbf{86.9}\\
MAP@3   & 75.4 & 87.5 & 86.6 & 87.6 & \textbf{88.0}\\
MAP@5   & 74.5 & 87.1 & 85.9 & 87.0 & \textbf{87.3}\\
MAP@10  & 72.5 & 85.9 & 85.0 & 85.6 & \textbf{86.0}\\
\hline
NDCG@1  & 73.4 & 86.3 & 84.9 & 86.5 & \textbf{86.9}\\
NDCG@3  & 68.3 & \textbf{84.2} & 83.0 & \textbf{84.2} & \textbf{84.2}\\
NDCG@5  & 65.6 & \textbf{82.4} & 81.7 & 82.1 & \textbf{82.4}\\
NDCG@10 & 58.7 & \textbf{79.1} & 78.0 & 77.8 & 78.6\\
\hline
MRR     & 77.5 & 88.4 & 87.3 & 88.5 & \textbf{88.7}\\
\hline
\hline
\end{tabular}
\end{table}

\begin{table}[t]
\centering
\caption{{\it Segment-based speaker retrieval performance on Q$^\dagger$.}}
\label{tab:seg-vs-spk}
\begin{tabular}{l|c|c|c|c}
\hline \\[-3mm]
Embedding & P@1 & P@3 & P@5 & P@10 \\
\hline
\hline \\[-3mm]
x-vector-SB     & 72.5 & 67.0 & 64.2 & 56.0 \\
ECAPA-TDNN-SB   & 86.5 & 83.4 & 81.2 & 76.7 \\
ECAPA-TDNN-NeMo & 84.3 & 82.2 & 80.4 & 75.3 \\
TitaNet-S-NeMo  & 86.8 & 83.7 & 81.1 & 75.8 \\
TitaNet-L-NeMo  & 86.4 & 83.3 & 81.2 & 76.5 \\
\hline
\hline
\end{tabular}
\end{table}

\subsection{Comparison of Speaker Embeddings}
\label{subsec:comp}
Table~\ref{tab:metrics} presents the performance of various speaker embeddings on the Q$^\dagger$ query set, using different retrieval performance metrics. Notably, the x-vector exhibits the poorest performance compared to other embeddings. The ECAPA-TDNN model from SpeechBrain consistently outperforms its counterpart from the NeMo toolkit. Additionally, we conducted a comparison between TitaNet-S and TitaNet-L, with the larger model demonstrating superior performance. Overall, TitaNet-L-NeMo and ECAPA-TDNN-SB yield the highest performance among the evaluated embedding types.

Note that despite being trained on a larger volume of data, NeMo models do not remarkably outperform SpeechBrain due to differences in the quality and relevance of the datasets used. Both were trained on VoxCeleb, a database specifically designed for speaker recognition in the wild, capturing a wide range of acoustic conditions representative of real-world scenarios. 
However, the additional datasets used by NeMo were primarily recorded in controlled, laboratory-like environments and were originally designed for other tasks such as speech recognition. These datasets lack the variability and complexity of real-world acoustic conditions, such as those found in the Rewind corpus. Consequently, the sheer quantity of training data does not necessarily translate into performance gains for speaker retrieval in uncontrolled and noisy environments. 

Table~\ref{tab:seg-vs-spk} reports the retrieval performance of the segment-level embeddings. Comparing the results with the speaker-level embeddings in Table~\ref{tab:metrics} shows that both segment-level and speaker-level embeddings demonstrate largely similar performance in the speaker retrieval task. 
Such a similarity is important because it indicates that the embeddings are primarily sensitive to the speaker information while remaining invariant to other information encoded in speech such as the lingual content which differs greatly across various segments. 
It is worth noting that the speaker-based retrieval is computationally more efficient than the segment-based retrieval, as discussed in Section~\ref{sec:stats}.

We also explored the usefulness of combining the top-performing embeddings, namely TitaNet-L-NeMo and ECAPA-TDNN-SB. To this end, before ranking the files based on the cosine similarity score, we combined the scores of each archive file as follows:
\begin{align}
\textit{Score} &= \lambda \ \textit{Score}_{\textit{ECAPA-TDNN}} \ + \ (1-\lambda) \ \textit{Score}_{\textit{TitaNet-L}}
\end{align}
\noindent where $\lambda$ is an interpolation parameter.

\begin{figure}[t!]
\centering
  \includegraphics[width=0.95\linewidth]{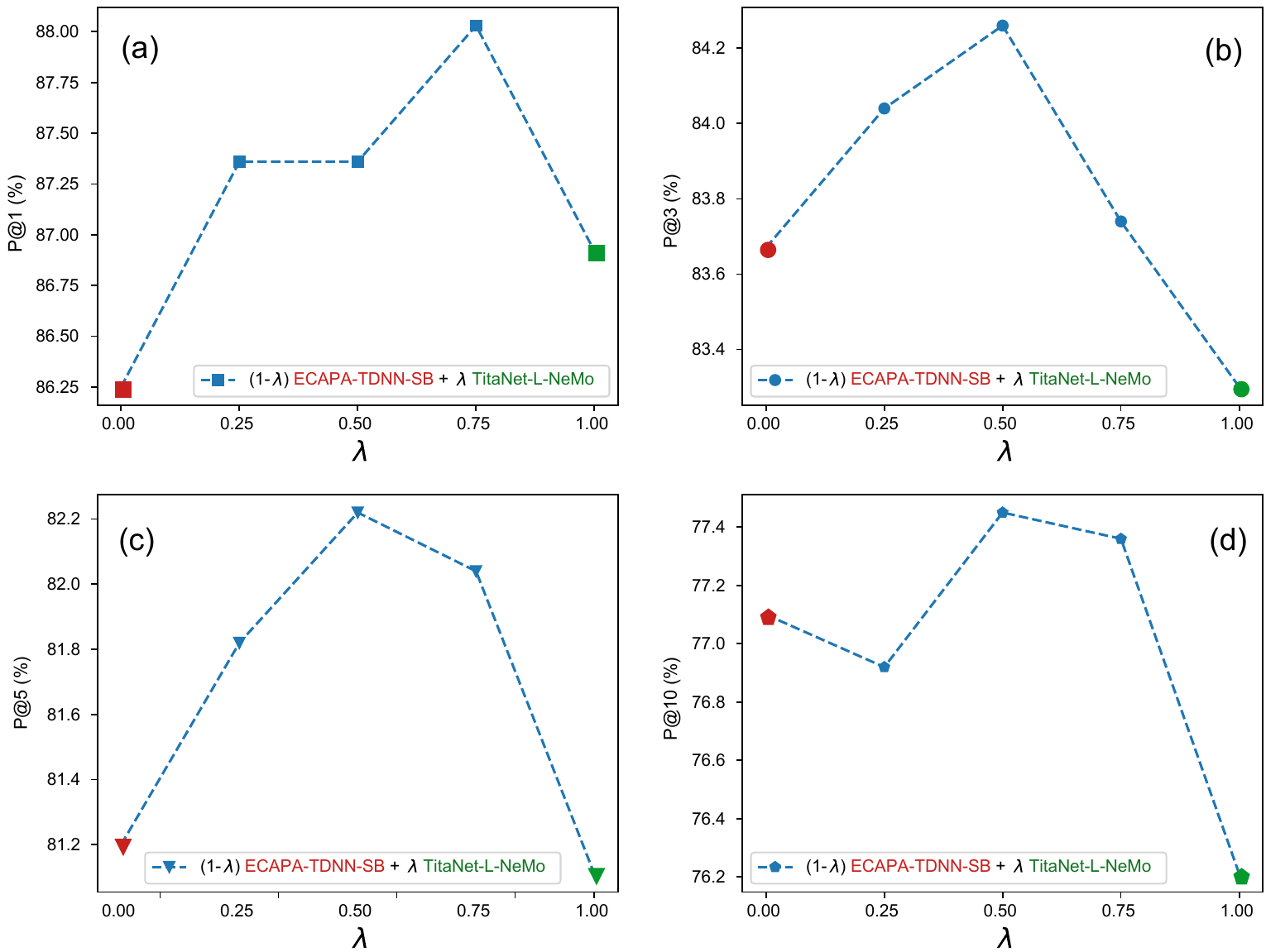}
  \caption{Effect of ECAPA-TDNN and TitaNet-L speaker embeddings combination on the retrieval performance. (a) P@1, (b) P@3, (c) P@5, (d) P@10.}
  \label{fig:fusion}
\end{figure}

Fig.~\ref{fig:fusion} depicts that combining the ECAPA-TDNN-SB and TitaNet-L yields a small yet consistent performance enhancement, especially when $\lambda$ is optimised. As seen, $\lambda=0.5$ yields the most consistent improvements. This performance improvement indicates that these models can synergise effectively, due to their diverse training data and distinct architectures, enabling complementary modeling approaches.

\begin{figure}[t!]
\centering
  \includegraphics[width=\linewidth, height=80mm]{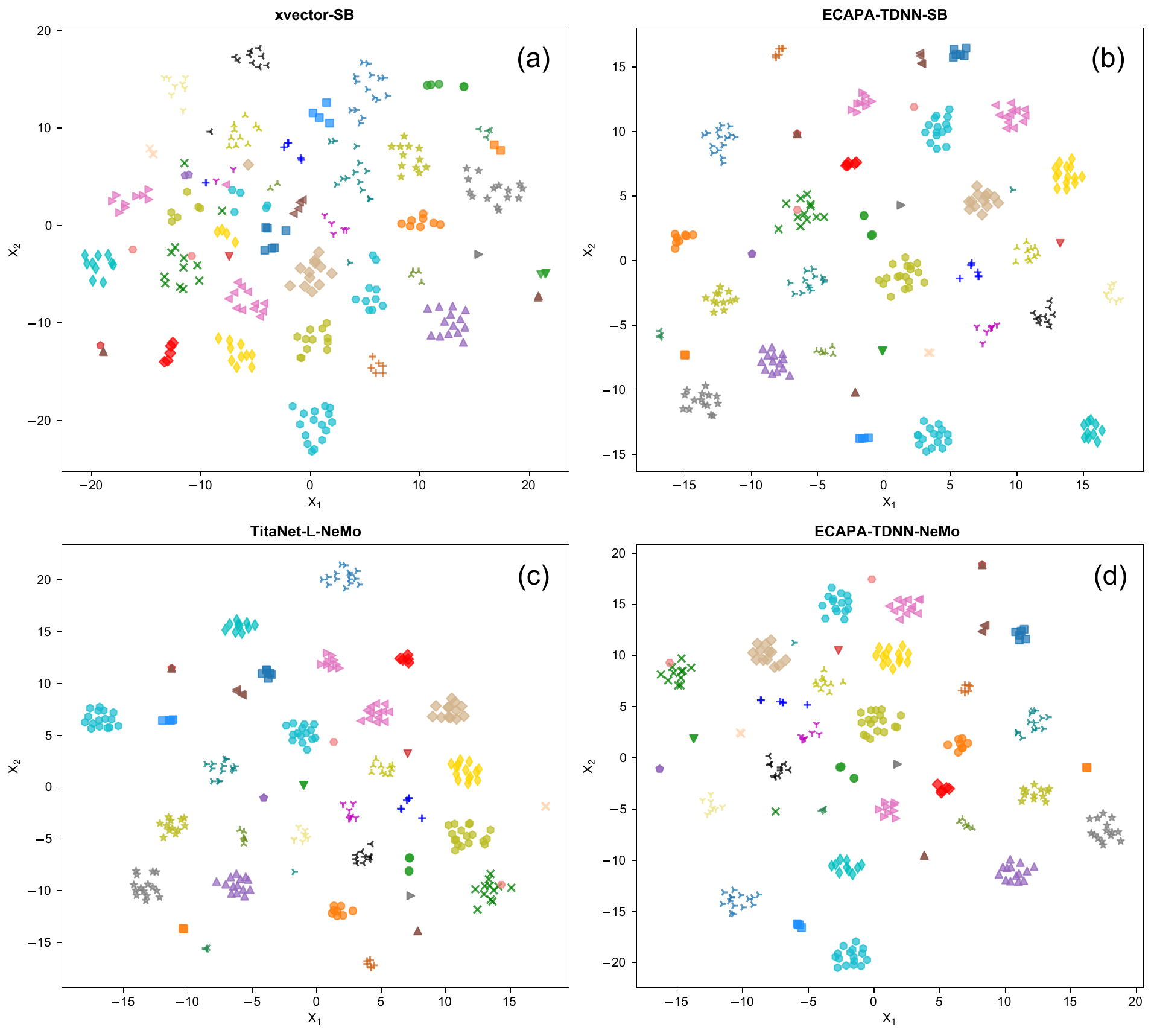}
  \caption{Scatter plots in 2D using t-SNE dimensionality reduction method. (a) x-vector, (b) ECAPA-TDNN-SB, (c) TitaNet-L-NeMo, (d) ECAPA-TDNN-NeMo. Each marker and color combination represents a unique speaker.}
  \label{fig:tsne-2d}
\end{figure}

Fig.~\ref{fig:tsne-2d} presents a scatter plot in 2D, generated after dimensionality reduction using the t-SNE method \cite{van2008tsne,scikit-learn}, for 300 out of 451 query files of Q$^\dagger$. On these 300 query files, all the speaker embedding extraction methods achieved a perfect P@1. The aim of applying this query subset and visualisation is to assess the quality and separation of clusters formed by various embeddings while ensuring that no embedding extraction method is disadvantaged for its lower performance. 

Upon visual inspection, it is observed that the clusters formed by x-vector are less distinct compared to others while TitaNet-L leads to the best-formed clusters. 
To complement this qualitative evaluation, the \textit{Silhouette} score \cite{rousseeuw1987silhouettes,scikit-learn} was employed to quantify clustering quality. The Silhouette score, ranging from -1 to 1, with higher values indicating better-defined clusters.
Table~\ref{tab:sil} presents the results, revealing that x-vector and TitaNet-L achieve the lowest (0.333) and highest (0.709) Silhouette scores, respectively. This is consistent with the observations from Fig.~\ref{fig:tsne-2d}.

Overall, these experiments demonstrate the effectiveness of the developed speaker retrieval framework, constructed by leveraging pre-trained models. It is noteworthy that despite neither training nor fine-tuning these models on the target database and in spite of encountering a wide range of acoustic conditions across this archive, the developed speaker retrieval systems exhibit strong performance.

In the next section, we investigate the robustness of the developed system in a noisy setup, simulating the real-world applications, beyond the BBC Rewind corpus.

\begin{table}[t]
\centering
\caption{{\it Silhouette score for different speaker embeddings, showing the quality of clustering. XV: x-vector, ET: ECAPA-TDNN, TL: TitaNet-Large (TitaNet-L).}}
\label{tab:sil}
\begin{tabular}{l|c|c|c|c}
\hline \\[-3mm]
        & XV-SB & ET-SB & ET-NeMo & TL-NeMo \\
\hline   
\hline
Silhouette Score & 0.333 & 0.702 & 0.679 & \textbf{0.709} \\
\hline   
\hline
\end{tabular}
\end{table}

\section{Exploring the Robustness}
\label{sec:robustness}
To evaluate the robustness of the speaker retrieval system, we systematically introduced synthetic distortions to the query files while keeping the archive unchanged. These distortions aimed to simulate challenging scenarios that the system might encounter in various real-world applications. Specifically, we examined the system's robustness across a range of adverse acoustic conditions, including additive background noise, bit depth reduction, sampling rate reduction, and reverberation. 

In this series of experiments, we use Q$^\dagger$ along with the top-performing embeddings, namely ECAPA-TDNN-SB and TitaNet-L-NeMo. For performance evaluation, and to save space, we will utilise P@\{1, 3, 5, 10\}.

\subsection{Robustness against Additive Noise}
To explore the robustness of the speaker retrieval system against additive background noise, the query files were contaminated with four types of noise: Babble, White Gaussian (WG), Street, and Music (Jazz). Each noise type was added at SNRs of 15, 10, 5, and 0 dB. 

\begin{figure}[t!]
\centering
  \includegraphics[width=\linewidth, height=38mm]{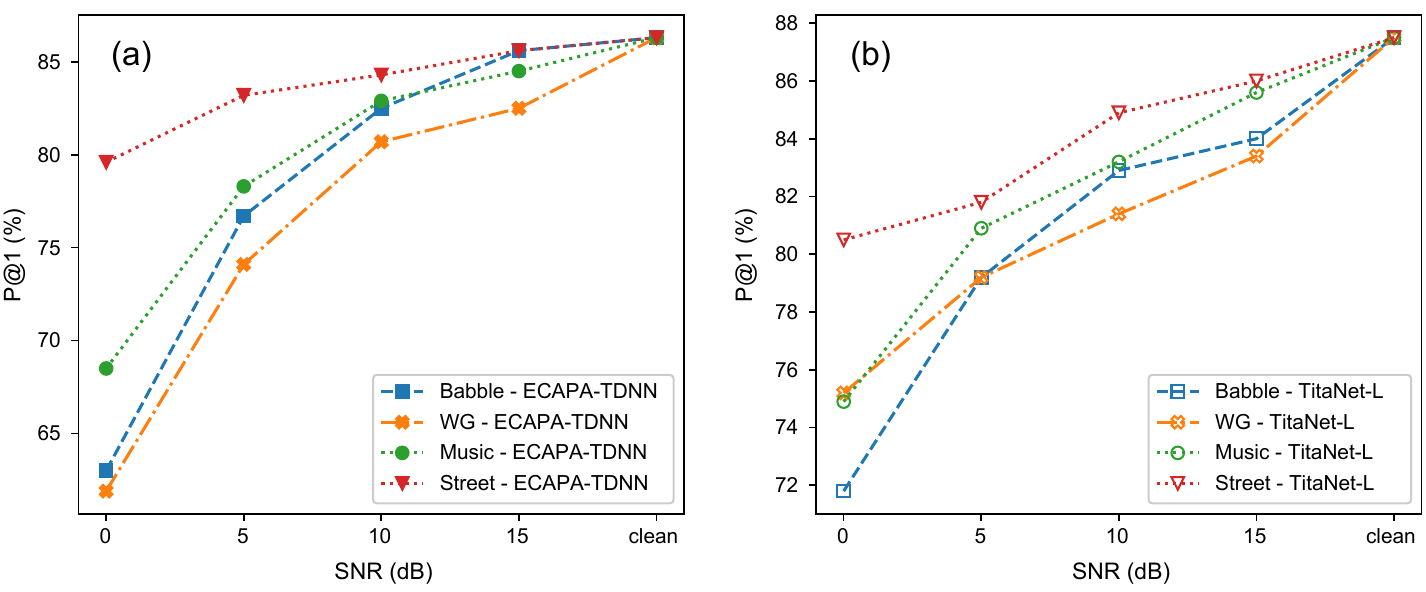}
  \caption{Retrieval performance for Babble, White Gaussian (WG), Street and Music noise, at different SNRs. (a) ECAPA-TDNN-SB, (b) TitaNet-L-NeMo.}
  \label{fig:all-noise}
\end{figure}

Fig.~\ref{fig:all-noise} depicts the P@1 vs SNR for the ECAPA-TDNN-SB and TitaNet-L-NeMo embeddings. Observing the results, the Babble and WG noise pose greater challenges while the Street noise falls in the middle, and the Music noise emerges as the least challenging among the tested noise types. 
The Babble noise's similarity to human vocalisation complicates its disentanglement from speech. Consequently, its undesirable effects are likely to propagate to higher levels in the embedding extraction model, leading to more significant distortions in the embedding space. 
Conversely, music noise is categorically distinct from speech, and therefore, is easier to disentangle and discard along the embedding extraction process. 

\begin{figure}[t!]
\centering
  \includegraphics[width=\linewidth, height=76mm]{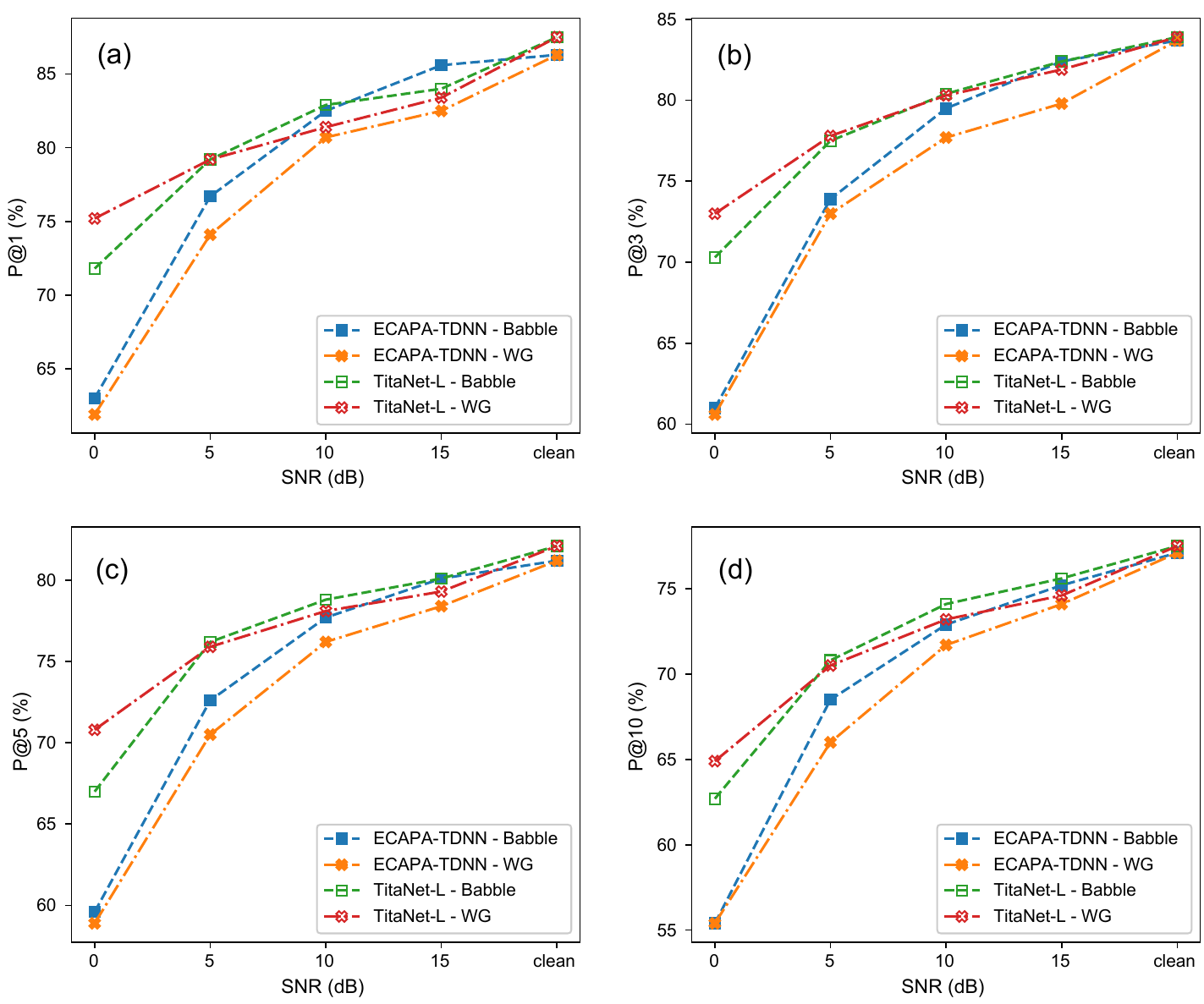}
  \caption{Speaker retrieval performance at different SNRs for different embeddings and noise types. (a) P@1, (b) P@3, (c) P@5, (d) P@10.}
  \label{fig:babble-wg-noise}
\end{figure}

Fig.~\ref{fig:babble-wg-noise} compares the performance of the ECAPA-TDNN-SB and TitNet-L-NeMo in terms of P@\{1,3,5,10\} on the Babble and WG noise at different SNRs. For SNRs larger than 10 dB, both models lead to similar performance. However, the TitaNet-L-NeMo outperforms ECAPA-TDNN-SB in SNRs below 10 dB, and the performance gap between them becomes notable in SNRs of 5 and 0 dB.

It is important to note that in an archive with over 5,800 speakers (none of whom were seen during training), even slight perturbations in the embedding space have the potential to shift the embeddings to neighborhoods that could mislead the retrieval process. The demonstrated performance highlights considerable noise robustness, particularly for TitaNet-L.


\subsection{Robustness against bit depth}
Reducing the bit depth at the waveform level increases the quantisation error. 
Robustness to bit depth reduction is valuable in applications where speech data is stored with reduced precision or transmitted over telephone networks, voice-over-IP (VoIP), and low-bitrate audio streaming.

Table~\ref{tab:bit} displays the performance of the speaker retrieval systems when the bit depth is reduced from 16 to 8 bits.  
Notably, at the 8-bit depth, both systems maintain high performance levels, with an average relative performance reduction (AvgRPR) of around -2\%. This highlights the resilience of these systems to moderate reductions in bit depth. The slightly lower average performance drop for the TitaNet-L model can be attributed to the fact that NeMo models are trained on Fisher and Switchboard data with an original 8-bit resolution.

\begin{table}[t]
\centering
\caption{{\it Speaker retrieval robustness against bit depth of 8 bits.}}
\label{tab:bit}
\begin{tabular}{l|c|c|c|c|c}
\hline \\[-3mm]
  Embedding & P@1 & P@3 & P@5 & P@10 & AvgRPR\\
\hline
\hline \\[-3mm]
ECAPA-TDNN-SB & 83.4 & 81.6 & 80.0 & 75.7 & -2.3 \\
\hline
TitaNet-L-NeMo  & 84.7 & 82.0 & 80.0 & 75.5 & -1.6 \\
\hline
\hline
\end{tabular}
\end{table}

\subsection{Robustness against Sampling Rate}
Robustness to sampling rate variations is essential for reliable speaker retrieval across various applications, e.g., in telecommunication and speech transmission over telephone networks, where bandwidth limitations often necessitate reduced sampling rates. In forensic audio analysis, evidence sources may have varied sampling rates, requiring the system to handle this variability effectively. Additionally, 
in voice-controlled devices or voice assistants, where audio signals may be captured at lower sampling rates for efficiency, resilience to this type of distortion is desired. 

We evaluated the robustness to sampling rate reduction through first downsampling the query set from 16 kHz to 8 kHz and to 4 kHz (decimation by factors of 2 and 4, respectively) and subsequently upsampling to restore the original 16 kHz sampling rate, as all pre-trained models operate at 16 kHz. This process simulates the loss of frequencies above 4 kHz and 2 kHz, respectively. 

The performance under 8 kHz (decimation factor: 2) and 4 kHz (decimation factor: 4) sampling rates is reported in Table~\ref{tab:sampling}. As the sampling rate decreases, there is a corresponding decline in performance. For both ECAPA-TDNN-SB and TitaNet-L-NeMo, the decrease in performance (relative to baseline) for the 8 kHz system is modest, at around 7.5\%. However, a remarkable drop in performance is observed when sampling at 4 kHz. These observations indicate the importance of spectral components between 2 to 4 kHz in the task of speaker identification and retrieval.
Additionally, we observe that the performance drop for the TitaNet-L-NeMo model is slightly lower than that for the ECAPA-TDNN-SB model. This difference can be attributed to the fact that the NeMo models are trained with Switchboard and Fisher data with sampling rates of 8 kHz, potentially providing them with better adaptability to lower sampling rates.


\begin{table}[t]
\centering
\caption{{\it Speaker retrieval robustness against variation in sampling rate (SR).}}
\label{tab:sampling}
\begin{tabular}{l|c|c|c|c|c|c}
\hline \\[-3mm]
  Embedding & SR (kHz) & P@1 & P@3 & P@5 & P@10 & AvgRPR \\
\hline
\hline \\[-3mm]
ECAPA-SB & 8 & 80.9 & 77.3 & 74.8 & 69.8 & -7.8 \\
ECAPA-SB & 4 & 36.8 & 32.6 & 31.3 & 28.6 & -60.7 \\
\hline
TitaNet-L & 8 & 78.3 & 78.0 & 75.5 & 71.2 & -7.4 \\
TitaNet-L & 4 & 42.6 & 41.0 & 39.5 & 36.0 & -51.5 \\
\hline
\hline
\end{tabular}
\end{table}

\subsection{Robustness against Reverberation}
The robustness against reverberation was evaluated using room impulse responses (RIRs) obtained from the Aachen Impulse Response (AIR) database \cite{jeub09a,jeub10a} and synthetic RIRs generated with the \textit{RIR-Generator} package \cite{rir-gen,habets2006room}. The AIR database offers diverse room configurations with different reverberation time (RT60) including Office (RT60=0.48 sec), Lecture room (RT60=0.79 sec), and the highly-reverberant Aula Carolina Church (RT60 not specified). 
The synthetic RIRs were generated using the image method \cite{allen1979image}, with varying RT60 from 0.25 to 2 seconds while keeping other room parameters at default values \footnote{RIR-Generator default setting: c=340, fs=16,000, s=[2, 3.5, 2], L=[5,4,6], microphone\_type=omnidirectional, r=[[2,1.5,1], [2,1.5,2],[2,1.5,3]]}. Each RIR was convolved with the original queries before extracting the speaker embeddings. 

Note that handling reverberation presents a greater challenge compared to other forms of noise and distortion studied here. Reverberation closely resembles speech characteristics and is highly non-stationary, making it difficult to disentangle its negative effects at the speaker embedding space. The higher the RT60, the more challenging the reverberation becomes.

Tables~\ref{tab:air-rir} and \ref{tab:gen-rir} present the speaker retrieval performance across different conditions.
Despite the challenges posed by reverberation, the ECAPA-TDNN-SB system maintains robust performance even in moderately reverberant environments, such as Office or Lecture room, or with RIRs having RT60 $<$ 1.0 sec, as evidenced in Tables~\ref{tab:air-rir} and \ref{tab:gen-rir}. However, the TitaNet-L-NeMo fails dramatically in dealing with reverberation when RT60 $>$ 0.25. This observation is rather surprising as this model is trained with RIR noise database, warranting further investigation by the NeMo development team.

\begin{table}[t]
\centering
\caption{{\it Speaker retrieval robustness against reverberation. Database: AIR.}}
\label{tab:air-rir}
\begin{tabular}{l|c|c|c|c|c}
\hline \\[-3mm]
  Embedding & Room  & P@1 & P@3 & P@5 & P@10 \\
\hline
\hline \\[-3mm]
ECAPA-SB & Office   & 83.4 & 80.5 & 77.9 & 73.3 \\
ECAPA-SB & Lecture  & 81.6 & 78.0 & 75.4 & 71.2 \\
ECAPA-SB & Aula Carolina & 75.4 & 72.4 & 71.1 & 67.0 \\
\hline
TitaNet-L-NeMo & Office & 83.4 & 80.9 & 78.7 & 74.5 \\
TitaNet-L-NeMo & Lecture & 55.2 & 56.8 & 57.4 & 55.6 \\
TitaNet-L-NeMo & Aula Carolina  & 63.6 & 63.6 & 63.4 & 60.3 \\
\hline
\hline
\end{tabular}
\end{table}

\begin{table}[t]
\centering
\caption{{\it Speaker retrieval robustness against reverberation. RIRs synthetically generated by the RIR-Generator package \cite{rir-gen} using the image method \cite{allen1979image}.}}
\label{tab:gen-rir}
\begin{tabular}{l|c|c|c|c|c|c}
\hline \\[-3mm]
  Embedding & RT60 & P@1 & P@3 & P@5 & P@10 & AvgRPR\\
\hline
\hline \\[-3mm]
ECAPA-SB & 0.25 & 84.9 & 82.5 & 80.0 & 75.5 & -1.7 \\
ECAPA-SB & 0.5  & 80.3 & 78.3 & 76.4 & 71.4 & -6.7 \\
ECAPA-SB & 1.0  & 72.1 & 68.4 & 66.3 & 62.7 & -17.9\\
ECAPA-SB & 1.5  & 57.0 & 55.5 & 54.1 & 50.6 & -33.9\\
ECAPA-SB & 2.0  & 33.3 & 33.0 & 32.0 & 30.9 & -60.6\\
\hline
TitaNet-L-NeMo & 0.25 & 82.7 & 80.3 & 78.5 & 73.5 & -3.8\\
TitaNet-L-NeMo & 0.5  & 69.8 & 69.3 & 68.0 & 64.7 & -16.9\\
TitaNet-L-NeMo & 1.0  & 18.6 & 18.9 & 20.9 & 24.6 & -74.5\\
TitaNet-L-NeMo & 1.5  & 7.5 & 7.2 & 7.7 & 9.1 & -90.3 \\
TitaNet-L-NeMo & 2.0  & 3.8 & 3.9 & 4.3 & 4.9 & -94.8 \\
\hline
\hline
\end{tabular}
\end{table}

\section{Conclusion and Scopes for Future Work}
In this study, we investigated the challenges associated with speaker retrieval in real-world scenarios, with a particular focus on the complexities of extensive and aged media archives, exemplified by the publicly available BBC Archive Rewind corpus. 
Our approach to system development ``in the wild'' encompasses dealing with two key challenges: first, extraction of the task-relevant reliable labels from limited metadata for the development and evaluation purposes; and second, the need to address a broad spectrum of acoustic conditions within the archive data, ranging from clean to very noisy environments.
To tackle these challenges, we employed the state-of-the-art ECAPA-TDNN and TitaNet speaker embedding extraction models, pre-trained on over 2,000 and 3,000 hours of data, respectively.
Through extensive experimentation, we showcased the effectiveness of our systems and demonstrated their robustness across a range of distortions, including various background noise types (Babble, Street, Music, and White Gaussian), different bit depths (16 and 8 bits), varied sampling rates (16, 8, and 4 kHz), and reverberation (utilising both real and synthetic room impulse responses).
These systematic experiments not only confirm the systems' versatility in handling dynamic adverse acoustic conditions but also establish their potential for reliable and accurate speaker retrieval in real-world applications beyond the BBC Rewind corpus.

While our developed speaker retrieval systems demonstrate promising capabilities, there are ample opportunities for future research. Our experiments have pinpointed some weaknesses in the deployed models, offering insights for addressing these issues to enhance their performance.
Exploring techniques to mitigate label ambiguity during training, such as incorporating semi-supervised learning, could further improve the system's performance by enabling fine-tuning on the target dataset. Finally, exploring the integration of additional modalities and the development of multimodal retrieval systems, represents another appealing and broad avenue for future research.

\bibliographystyle{IEEEtran}
\bibliography{ref}

\end{document}